\documentclass[twocolumn,aps,prb,amsmath,floatfix]{revtex4}
\usepackage{graphicx}
\begin{document}

\title{Magnetization dynamics in dysprosium orthoferrites via inverse Faraday 
effect}
\author{C.A. Perroni and A. Liebsch}
\affiliation{Institut f\"ur Festk\"orperforschung (IFF),
         Forschungszentrum J\"ulich,
         52425 J\"ulich, Germany}

\begin{abstract}
The ultrafast non-thermal control of magnetization has recently become 
feasible in canted antiferromagnets through photomagnetic instantaneous
pulses [A.V. Kimel {\it et al.}, Nature {\bf 435}, 655 (2005)]. In this experiment
circularly polarized femtosecond laser pulses set up a strong magnetic
field along the wave vector of the radiation through the inverse Faraday effect, 
thereby exciting non-thermally the spin dynamics of dysprosium orthoferrites. 
A theoretical study  is performed by using a model for orthoferrites based on a general 
form of free energy whose parameters are extracted from experimental measurements. The 
magnetization dynamics is described by solving coupled sublattice 
Landau-Lifshitz-Gilbert equations whose damping 
term is associated with the scattering rate due to magnon-magnon interaction. Due 
to the inverse Faraday effect and the non-thermal excitation, the 
effect of the laser is simulated by magnetic field Gaussian pulses with 
temporal width of the order of hundred femtoseconds. When the field is along 
the z-axis, a single resonance mode of the magnetization is excited. 
The amplitude of the magnetization and out-of-phase behavior of the oscillations 
for fields in z and -z directions are in good agreement with 
the cited experiment. The analysis of the effect of the temperature 
shows that magnon-magnon scattering mechanism affects the decay of the 
oscillations on the picosecond scale. Finally, when the field 
pulse is along the x-axis, another mode is excited, as observed in experiments.
In this case the comparison between theoretical and experimental results   
shows some discrepancies whose origin is related to the role played by anisotropies 
in orthoferrites.  
\end{abstract}
\maketitle

\section{Introduction}
In the last years  the need of enhancing the speed of modern spin-electronic and 
magneto-optic devices, and of further developing the magnetic storage technology, has 
stimulated many studies aimed at achieving a fundamental understanding of the 
mechanisms of magnetization dynamics and switching. In addition to ultrafast magnetic
field pulses which require complex devices for their generation, the spin dynamics has 
been induced by ultrafast optical laser pulses. Recent experiments have shown that 
significant demagnetization of magnetic compounds can be measured on the scale of a few 
hundred femtoseconds. \cite{beaure,hohle,schol,koopm}   
Typically the light is absorbed by the material, giving rise to a rapid increase of
temperature responsible for the change of magnetization \cite{hubne,guido,ju} and spin 
reorientation. \cite{kimel} The question concerning the exact speed of the initial 
sub-picosecond magnetization breakdown is still subject of debate and partly relates to 
the question of how to interpret magneto-optical experiments. 
\cite{koopm,koopm1,liebs} Furthermore, the cooling time could limit the repetition 
frequency whose value is fundamental for actual applications. \cite{hohle1}   

Recently, non-thermal ultrafast optical control of magnetization has been achieved 
in canted antiferromagnet samples of dysprosium orthoferrites by using circularly 
polarized femtosecond pulses. \cite{kimel1} Via the inverse Faraday effect, the light 
excitation acts on the spins of the system as a magnetic field pulse directed along 
the wave vector of the radiation and proportional to its intensity. \cite{ziel,pershan} 
The inverse Faraday effect does not rely on absorption and has its fingerprint in the 
fact that the helicity of the pump controls the sign of the photo-induced 
magnetization.  This effect plays a role also in the femtosecond photomagnetic 
switching of spins in ferromagnetic garnet films.\cite{hanst}

Since the manipulation of spins by means of circularly polarized laser 
pulses represents an advance in the field of ultrafast magnetization dynamics, 
we analyze thoroughly the experimental work by Kimel {\it et al.} \cite{kimel1}
and discuss its pecularities. In this experiment the difference between the Faraday 
rotations induced by right- and left-handed polarized pulses has been studied in the 
temperature range between 20 K and 175 K. A characteristic spin-wave mode, called 
quasi-antiferro mode, is excited by the light pulse along the z-direction.
When the sample is heated, the frequency of the mode oscillations increases and 
the amplitude decreases from the low-temperature maximum value of the order of $M_S/16$,
where $M_S$ is the saturation magnetization. We point out that there is no 
theoretical explanation concerning the magnitude and the temperature behavior of the 
oscillations. Furthermore, some aspects of the cited experiment deserve attention. 
Actually, in the so called $\Gamma_4$ phase, stable at temperatures higher than 50 K, 
the oscillations have temperature-dependent frequencies in full agreement with 
those measured by Raman experiments. \cite{white,koshi} On the other hand,  
at temperatures below 50 K, the frequency of the photoinduced magnetization 
stays constant in contrast with the results of the Raman spectra and 
other measurements which signal the discontinuous transition to another phase, 
called $\Gamma_1$. Therefore, the experimental data obtained by laser pulses 
reflect the excitation of resonance modes characteristic of the $\Gamma_4$ phase 
even at very low temperatures. Finally, the field pulse is also directed along the 
x-axis: another spin-wave mode, called quasi-ferro mode, is excited and 
is characterized by an extraordinarily, not well understood, small amplitude. 

In order to obtain a deeper understanding of the non-thermal control of magnetization, 
in this paper we perform simulations related to the experiment by 
Kimel {\it et al.} \cite{kimel1} We have studied the 
magnetization dynamics employing a model for orthoferrites that was previously proposed 
for the analysis of resonance and high-frequency susceptibility. \cite{hermann} 
The parameters of the free energy, such as the symmetric and antisymmetric exchange, and
the anisotropy constants, are determined by using the experimental Raman spectra of 
Ref.17. The dynamical behavior is described by solving two coupled sublattice nonlinear
Landau-Lifshitz-Gilbert equations through a fifth-order Runge-Kutta algorithm. 
The damping term in the dynamical equations is related to magnon-magnon interaction 
and its temperature behavior is provided by a calculation of the scattering rate in 
orthoferrites. \cite{tsang}  
Exploiting the inverse Faraday effect, we have analyzed the effect of Gaussian 
magnetic field pulses whose time width is of the order of hundred femtoseconds. 
Since, in the regime considered in the experiments, the effective magnetic fields are 
not large if compared with exchange fields, the solution of the linearized system 
represents a reasonable approximation to the numerical results. Therefore 
we have studied the dynamics within the linear solution after the excitation by a pulse 
shaped as a delta function, since the magnetic field pulse takes place on a time scale 
shorter than the period of the resonance modes.

One result of this work is that, in the $\Gamma_4$ phase, the quasi-antiferro mode of
the magnetization is excited by a field pulse along the z-axis and the 
oscillations have amplitudes in agreement with experimental results. 
Moreover, the oscillations induced by pulses directed along the z and -z axis show the 
characteristic out-of-phase behavior.  
We point out that, even for the dynamics, the ratio between the antisymmetric and 
symmetric exchange energies is important. Furthermore, we stress that in the 
canted antiferromagnets, such as rare-earth orthoferrites, the largest amplitudes of the
oscillation are not obtained for the ferromagnetic sum vector of the sublattice 
magnetizations but for the antiferromagnetic difference vector. The behavior of the 
magnetization has been analyzed in the temperature range between 20 K and 175 K. The 
damping process based on magnon-magnon scattering describes the 
decay of the oscillations with results consistent with the experiment. 

The case of the field pulse directed along the x-axis has been analyzed in the 
$\Gamma_4$ phase. The magnetization along the x-axis oscillates with the frequency of 
the quasi-ferro mode as found in the experiment. However, the calculated and the 
experimental amplitudes of the oscillation are different if the magnetic field pulse 
along the x-axis has the same intensity of that along the z-axis. 
Therefore, in the comparison with experimental data, we discuss the role
of anisotropies between z and x directions as a source of discrepancy between 
theory and experiment. 

Finally, we consider the actual stable phase in equilibrium at low temperatures, 
the $\Gamma_1$ phase, and its resonance modes. We point 
out that the difference in energy between the $\Gamma_1$ and $\Gamma_4$ can be very 
small, so that even a small laser-heating effect could be responsible for the 
stabilization of the $\Gamma_4$ phase at very low temperatures on a picosecond time 
scale.

The outline of this paper is as follows. In the next section we discuss
the numerical approach for the system and the analytic 
solution of the linearized equations for excitation by a delta function shaped 
magnetic field pulse. Section III provides the numerical and 
analytical results: in the first and second subsection the excitation due to the pulses 
along the z-axis and x-axis, respectively, is analyzed when the system at equilibrium 
is in the $\Gamma_4$ phase. In the final subsection the effect on the dynamics of a 
$\Gamma_1$ phase stable at low temperatures is considered. 
Section IV provides a summary.

\section{Free energy and dynamical equations}
Rare-earth orthoferrites are represented by the formula $ReFeO_3$, 
where $Re$ stands for rare-earth. They have a perovskite-type structure with 
slight deformation from cubic to orthorombic. In many of them the spins of iron 
ions are antiferromagnetically aligned through a strong super-exchange interaction 
with a N\`eel transition temperature of about 700K. Moreover, promoted by the 
orthorombic deformation, an antisymmetric exchange interaction acts between iron spins, 
resulting in spin canted magnetism with a feeble saturation moment. In the case of 
dysprosium orthoferrites, the temperature dependence of the ferromagnetic moment is 
characterized by a steep rise around 50 K in coincidence with the stabilization of the 
$\Gamma_4$ phase. \cite{wolf} Actually 
these compounds are also known for their spin reorientation properties: continuous 
rotational-type (ferromagnetism present in the low-temperature phase) and, only for 
dysprosium, abrupt-type from $\Gamma_1$ (antiferromagnetic) to $\Gamma_4$ with 
increasing temperature. 

\begin{figure}[t!]%1
  \begin{center}
  \includegraphics[width=6cm,height=7cm,angle=-90]{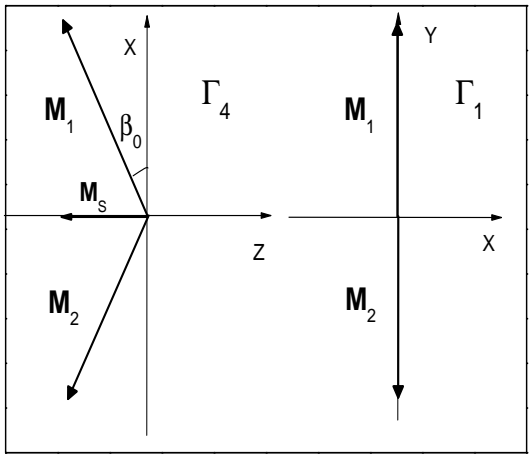}
  \end{center}
  \vskip-3mm
\caption{
The equilibrium positions of sublattice magnetization vectors for the high-temperature 
$\Gamma_4$ (ferromagnetic) and low-temperature $\Gamma_1$ (antiferromagnetic) phases. 
Due to the ratio between the antisymmetric $D$ and symmetric $E$ exchange fields, 
the angle $\beta_0$ is actually very small.
}\end{figure}

We use the free energy and the dynamical equations proposed in a previous work and 
focus on the behavior of the magnetization in the $\Gamma_4$ phase. \cite{hermann} 
The static and dynamical behavior is studied within the two-sublattice model for iron spins that takes into account the two active resonance modes. 
This represents an excellent approximation since the remaining resonance modes
are almost inactive, and their interaction with active modes is negligible. 
In Fig.1 we show a schematic representation of equilibrium positions of two-sublattice 
magnetization vectors for dysprosium orthoferrites. \cite{koshi}

The normalized free energy $V=F/M_0$, with $M_0$ the modulus of the sublattice 
magnetization, is composed of a part $V_{exc}$ due to  exchange interactions and a part 
$V_{ani}$ due to the anisotropy:
\begin{equation}
V= V_{exc}+V_{ani}.
\label{e0}
\end{equation}
The free energy is expanded as a power series in the magnetization components and, 
in order to properly describe the $\Gamma_4$ phase, only quadratic terms are sufficient. 
The exchange energy is written as the sum of a scalar and a pseudo-vector part
\begin{equation}
V_{exc}=E {\vec R}_1 \cdot {\vec R}_2 +D (X_1 Z_2-X_2 Z_1),
\label{e1a}
\end{equation}
where $E$ and $D$ are the symmetric and antisymmetric exchange fields, respectively, 
${\vec R}_1={\vec M}_1/M_0 \equiv (X_1,Y_1,Z_1)$, and 
${\vec R}_2={\vec M}_2/M_0 \equiv (X_2,Y_2,Z_2)$.
The anisotropic energy is
 \begin{equation}
V_{ani}=-A_{xx}(X_1^{2}+X_2^2)-A_{zz}(Z_1^2+Z_2^2).
\label{e1b}
\end{equation}        
The exchange field $E$ is quite large compared to the other terms. It is 
related to the exchange spin-spin interaction $J$ via the equation $E=12 J S / g \mu_B$, 
with $S=5/2$ the spin of iron ions, $g=2$, and $\mu_B$ the Bohr magneton. Since $J$ is of
the order of $20$ $cm^{-1}$, $E$ is approximately $6.4 \times 10^6$ Oe. \cite{koshi} 
The field $D$ is related to the antisymmetrical exchange energy $d$ through the relation
$D=6 d S / g \mu_B$ and is of the order of $1.4 \times 10^5$ Oe ($d$ is about 
$0.88$ $cm^{-1}$). The anisotropy constants $A_{xx}$ and $A_{zz}$ depend on 
temperature and are of the order of hundreds of Oersted. 
 
Using the energy given in Eq.(\ref{e0}), the equilibrium position can be derived for 
the $\Gamma_4$ phase. This is shown in Fig. 1, where the small canting angle $\beta_0$ 
is determined by the equation
\begin{equation}
tan(2 \beta_0)=\frac{D}{E+A_{xx}-A_{zz}} \simeq \frac{D}{E} =0.022
\label{e2}.
\end{equation}
Hence the magnetization  
$M_S=| \vec{M}_1+ \vec{M}_2 |=2 M_0 \sin (\beta_0)\sim 0.022 M_0$ is two orders of 
magnitude less than $M_0$.

The equilibrium position of $\vec{M}_1$ and $\vec{M}_2$ represents a stationary 
solution of the nonlinear Landau-Lifshitz-Gilbert equations 

\begin{equation}
\frac{1}{\gamma} \frac{d \vec R_1 }{dt}=-{\vec
R}_1 \wedge \left({\vec H}(t)-\vec{\nabla}_1 V \right)+\alpha
{\vec R}_1
\wedge \frac{d \vec R_1}{dt} 
\label{e3} 
\end{equation}
\begin{equation}
\frac{1}{\gamma} \frac{d \vec R_2}{dt}=-{\vec
R}_2 \wedge \left({\vec H}(t)-\vec{\nabla}_2 V \right)+\alpha
{\vec R}_2 \wedge \frac{d \vec R_2}{dt}, \label{e4}
\end{equation}
where $\gamma= 17.6$ MHz/Oe is the gyroscopic ratio, $\vec{\nabla}_1$ and 
$\vec{\nabla}_2$ are gradients with respect to $\vec R_1$ and $\vec R_2$, 
respectively, and $V$ is the energy in Eq.(\ref{e0}). 
Clearly the dynamical equations satisfy the following constraints: 
$X_1^2+Y_1^2+Z_1^2=1$ and $X_2^2+Y_2^2+Z_2^2=1$.

The quantity ${\vec H}(t)$ in Eqs.(\ref{e3},\ref{e4}) is the time-dependent magnetic 
field simulating the effect of laser pulses due to the inverse Faraday effect. In the 
experiment by Kimel {\it et al.} \cite{kimel1} the magnetic field pulse is of the 
order of fractions of Tesla with time width  of hundred femtoseconds. In the 
numerical simulations we consider a pulse directed along the propagation 
direction of the light with a Gaussian shape
\begin{equation}
{\vec H}(t)=\hat{k} \frac{F_0}{\sqrt{\pi} \tau_p} \exp\left[-(t/\tau_p)^2\right],
\label{e4a}
\end{equation}
where $\hat{k}$ defines the direction of the light wave-vector and $\tau_p $ indicates 
approximately the duration of the pulse. \cite{notarot}

Finally, in Eqs.(\ref{e3},\ref{e4}) $\alpha$ is the Gilbert constant. Actually, 
$\alpha$ takes into account the damping of the oscillations due to the magnon-magnon 
scattering and to the interaction of magnons with dysprosium spins and phonons. 
We notice that the scattering via dysprosium spins should be larger at very low 
temperatures where dysprosium ions tend to order. Moreover, the phonon-magnon 
scattering should be effective only on the nanosecond time scale. \cite{hanst,kimel2} 
It is the magnon-magnon interaction that provides the preminent source of 
scattering on the picosecond time scale.

\subsection{Solution of linearized system}
In this subsection we consider the solution determined by linearizing 
Eqs.(\ref{e3},\ref{e4}) and study the excitation of this linear system due to a 
magnetic field pulse shaped as a delta function. 
This is reasonable since, in the regime considered in the experiments, the effective 
magnetic fields are not large if compared with exchange fields, and the temporal width 
of the pulse is much smaller than the periods of the resonance modes.

In order to take into account small deviations from the equilibrium, the standard 
approach is to consider two separate coordinate systems, 
($S_1$,$T_1$,$Y_1$) and ($S_2$,$T_2$,$Y_2$), which describe the 
dynamics of  $\vec{M}_1$ and $\vec{M}_2$, respectively. \cite{hermann}
The variables $S_1$ and $S_2$ are chosen in order to coincide with the equilibrium 
positions of $\vec{M}_1$ and $\vec{M}_2$, respectively, so that
\begin{eqnarray}
S_1&=&\sin(\beta_0) Z_1+\cos(\beta_0) X_1, \\ 
T_1&=&- \cos(\beta_0) Z_1 + \sin(\beta_0) X_1,
\label{e4b}
\end{eqnarray}  
and
\begin{eqnarray}
S_2&=&\sin(\beta_0) Z_2-\cos(\beta_0) X_2, \\ 
T_1&=&\cos(\beta_0) Z_2 + \sin(\beta_0) X_2.
\label{e4c}
\end{eqnarray} 
By linearizing the system, we obtain the frequencies of two modes, the quasi-antiferro 
$\omega_{AFM}$ and the quasi-ferro $\omega_{FM}$ modes, involving cooperative motions 
of spins of the two sublattices. The energy of the quasi-antiferro and quasi-ferro  
modes is of the order of several $cm^{-1}$.

The first mode is characterized by the frequency 
\begin{equation}
\frac{\omega^2_{AFM}}{\gamma^2}=4 E A_{xx}+4A_{xx}(A_{xx}-A_{zz})+D^2.
\label{e5}
\end{equation}
The dynamic in this mode shows the following behavior: $\Delta X_1(t)=- \Delta X_2(t)$, 
$ \Delta Y_1(t)=- \Delta Y_2(t)$ and 
$ \Delta Z_1(t)=\Delta Z_2(t)$, where $ \Delta W_i(t) = W_i(t)-W_i^{eq}$, with $W=X,Y,Z$,
$i=1,2$, and $W_i^{eq}$ corresponding to the equilibrium position shown in Fig.1. 
Thus, the only ferromagnetic component different from zero is the magnetization along 
the z-axis with respect to the equilibrium,   $\Delta {M}_Z = M_0 \Delta \tilde{M}_Z$, 
where $\Delta \tilde{M}_Z$ is defined by 
\begin{equation}
\Delta \tilde{M}_Z = \Delta Z_1 (t)+\Delta Z_2 (t) = Z_1(t)+Z_2(t)-(Z_1^{eq}+Z_2^{eq}).
\label{e14a}
\end{equation}
Since the net spin does not reflect the motions of the sublattice spins, 
this mode is called quasi-antiferromagnetic. 

The second mode, with frequency
\begin{equation}
\frac{\omega^2_{FM}}{\gamma^2}=4E(A_{xx}-A_{zz})+4A_{xx}(A_{xx}-A_{zz}),
\label{e6}
\end{equation}
is characterized by $\Delta X_1(t)=\Delta X_2(t)$, $\Delta Y_1(t)=\Delta Y_2(t)$ and 
$\Delta Z_1(t)=-\Delta  Z_2(t)$. Since the net spin executes the
same rocking behavior as the sublattice spins, this mode is called quasi-ferromagnetic.

Eqs.(\ref{e5},\ref{e6}) relate the mode frequencies to the model parameters. The 
exchange fields are assumed constant in temperature at the values given above, since 
they represent the highest energy scales. Using Eqs.(\ref{e5},\ref{e6}) and the 
experimentally measured Raman spectra of Ref.17, we derive the anisotropy parameters. 
The quasi-antiferro mode increases linearly as a function of temperature in the 
$\Gamma_4$ phase: from 150 GHz at 50 K to 450 GHz at about 200 K, while the quasi-ferro 
mode stays constant at about 375 GHz. Therefore the anisotropic terms change upon 
heating the sample: $A_{xx}$ varies from about -640 Oe at T=50 K to about 200 Oe at 
T=200 K, $A_{zz}$ from -1540 Oe to -700 Oe in the same temperature range, with 
approximately fixed $A_{xx}-A_{zz}=900$ Oe. The knowledge of the frequencies is 
important also to determine the value of the damping constant in the dynamical 
equations 
(\ref{e3},\ref{e4}). 
Indeed, if $\omega_0$ is the temperature-dependent frequency of one of the modes, the 
damping constant $\alpha$ could be related to the damping rate $1/\tau_0$ by the 
relation $1/\tau_0=\omega_0 \alpha$.   
At the picosecond scale, the dominant spin-wave damping is due to four-magnon 
scattering. By using many-body perturbation theory, the rate for spin-wave at zero 
wave-vector is estimated to be $1/\tau_0 \sim 2.66 \cdot 10^4$ $T^2$ $s^{-1}$, with $T$ 
temperature in units of Kelvin degrees. \cite{tsang} The scattering rate grows 
quadratically in temperature, while the frequency of the quasi-antiferro mode is an 
increasing linear function of the temperature. \cite{koshi} The quantity 
$\alpha$ gets larger with temperature since the most important contribution is given by 
$1/\tau_0$. Actually, the values of $\alpha$ corresponding to 
the quasi-antiferro mode range from about $0.4 \cdot 10^{-4}$ at T=50 K to 
$3 \cdot 10^{-4}$ at T=200 K. The smallness of $\alpha$ implies that the oscillations 
of the magnetization are not strongly damped.      

It is useful to consider a magnetic field pulse shaped as a delta function
\begin{equation}
{\vec H}(t)=\hat{k} F_0 \delta(t).
\label{e6a}
\end{equation}
Since the pulse is instantaneous, it provides an initial condition to the dynamics 
described by the linearized equations of motions. We have analyzed two cases: 
field along z- and x-axis, since these are prominent for the experiment that we 
want to discuss. 

For the field along the z-axis, starting from the equilibrium position at $t=0^-$, we 
find at $t=0^+$
%\begin{eqnarray}
%X_1 &=& \cos(\beta_0) \cos(\gamma F_0)=-X_2 \\
%Y_1 &=& \cos(\beta_0) \sin(\gamma F_0) = - Y_2\\ 
%Z_1 &=& Z_1^{eq} = \sin(\beta_0) = Z_2 = Z_2^{eq}.  
%\label{e7}
%\end{eqnarray}  
%This implies that  
\begin{eqnarray}
\Delta X_1(0^+)&=&-\Delta X_2(0^+)=\cos(\beta_0) [\cos(\gamma F_0) -1] \\ 
\Delta Y_1(0^+)&=&-\Delta Y_2(0^+)= \cos(\beta_0) \sin(\gamma F_0) \\
\Delta Z_1(0^+)&=&\Delta  Z_2(0^+)=0.
\end{eqnarray}
Therefore this pulse will excite only the quasi-antiferro mode. 

The field along the x-axis yields at $t=0^+$
%\begin{eqnarray}
%X_1 &=& X_1^{eq} = \cos(\beta_0) = - X_2 = -X_2^{eq},\\ 
%Y_1 &=& - \sin(\beta_0) \sin(\gamma F_0)= Y_2, \\ 
%Z_1 &=& \sin(\beta_0) \cos(\gamma F_0) = Z_2.  
%\label{e9}
%\end{eqnarray}  

\begin{eqnarray}
\Delta X_1(0^+) &=&\Delta X_2(0^+) =0  \\ 
\Delta Y_1(0^+) &=&\Delta Y_2(0^+) =-\sin(\beta_0) \sin(\gamma F_0) \\
\Delta Z_1(0^+) &=&-\Delta  Z_2(0^+)= \sin(\beta_0) [\cos(\gamma F_0) -1].
\end{eqnarray} 
Thus the system will be excited in the quasi-ferro mode. 

As the experimental field magnitude is lower than the exchange fields, we study the
subsequent dynamical evolution within the linear approximation. \cite{hermann}
Within a relaxation time approximation, we include the damping time  $\tau$ 
describing the decay towards the equilibrium position. \cite{morrish}  

From the solution of the linear system in the case of a pulse along the z-axis, we get 
the evolution of the quasi-antiferro mode. The only ferromagnetic component, 
defined in Eq.(\ref{e14a}), is  
\begin{eqnarray}
\Delta \tilde{M}_Z=&&e^{- t / \tau} \frac{\cos(\beta_0)}{2} \sin(2 \beta_0) 
\sin^2(\frac{\gamma F_0}{2}) \cos(\omega_{AFM} t) \nonumber \\ 
&&+2 e^{- t / \tau}  R \cos^2(\beta_0) \sin(\gamma F_0)  
\sin(\omega_{AFM} t) \nonumber  \\
&&+2 e^{- t / \tau}   \sin{\beta_0} \cos^2(\beta_0) \left[\cos(\gamma F_0)-1 \right].
\label{e14}
\end{eqnarray}
We notice in Eq.(\ref{e14}) that the even terms in the magnetic field are at least two 
orders of magnitude less than the odd term. 

Along the other two directions, there are the antiferromagnetic component along x-axis 
(with respect to the equilibrium), $\Delta {A}_X = M_0 \Delta \tilde{A}_X$, with 
$\Delta \tilde{A}_X$ defined as
\begin{equation}
\Delta \tilde{A}_X=\Delta X_1(t)-\Delta X_2(t)=X_1(t)-X_2(t)-(X_1^{eq}-X_2^{eq}), 
\label{e15a}
\end{equation}
and the antiferromagnetic component along y-axis (with respect to the equilibrium), 
$\Delta {A}_Y = M_0 \Delta \tilde{A}_Y$, with  $\Delta \tilde{A}_Y$ 
\begin{equation}
\Delta \tilde{A}_Y=\Delta Y_1(t)-\Delta Y_2(t)=Y_1(t)-Y_2(t)-(Y_1^{eq}-Y_2^{eq}). 
\label{e16a}
\end{equation}
Using again the linearized form of Eqs.(\ref{e3},\ref{e4}), we obtain 
\begin{eqnarray}
\Delta \tilde{A}_X=&&-e^{- t / \tau} \sin(2 \beta_0) \sin(\beta_0) 
\sin^2(\frac{\gamma F_0}{2}) \cos(\omega_{AFM} t) \nonumber \\
&&- e^{- t / \tau}  R \sin(2 \beta_0) \sin(\gamma F_0)  
\sin(\omega_{AFM} t) \nonumber  \\
&&+2 e^{- t / \tau}   \cos{\beta_0} \cos^2(\beta_0) \left[\cos(\gamma F_0)-1 \right],
\label{e15}
\end{eqnarray}
and 
\begin{eqnarray}
\Delta \tilde{A}_Y=&&-e^{- t / \tau}  \frac{1}{R}  \sin(2 \beta_0) 
\sin^2(\frac{\gamma F_0}{2}) \sin(\omega_{AFM} t)  \nonumber \\ 
&&+2 e^{- t / \tau}  \cos(\beta_0) \sin(\gamma F_0) \cos(\omega_{AFM} t).
\label{e16}
\end{eqnarray}
The components specified in Eqs.(\ref{e14},\ref{e15},\ref{e16}) are 
determined by the mode frequency $\omega_{AFM}$, the angle $\beta_0$, and  
the quantity $R$ which depends on the model parameters in the form 
\begin{equation}
R=\sqrt{\frac{2 A_{xx}+D^2/2E}{2 E}}. 
\label{e16d}
\end{equation}
At intermediate temperatures ($A_{xx}$ very small) one gets $R \simeq D/2E$.

When the pulse is directed along the x-direction, the quasi-ferro mode is excited. 
The components of the magnetization different from zero are those along x- and y-axis. 
In particular we consider the magnetization along x-axis 
(with respect to equilibrium), $\Delta {M}_X = M_0 \Delta \tilde{M}_X$, where
$\Delta \tilde{M}_X$ is defined as
\begin{equation}
\Delta \tilde{M}_X=\Delta X_1(t)+\Delta X_2(t).
\label{e16c}
\end{equation}
From the linearized dynamical equations this quantity is calculated as
\begin{equation}
\Delta \tilde{M}_X= e^{- t / \tau}  \sqrt{\frac{4 E}{A_{xx}-A_{zz}}}  
\sin^2(\beta_0) \sin(\gamma F_0) \sin(\omega_{FM} t).
\label{e17}
\end{equation}

\section{Numerical and analytical results}
Using the procedure outlined in the previous section, all the parameters appearing
in the dynamical equations (\ref{e3},\ref{e4}) can be determined. These nonlinear 
equations are numerically integrated through a fifth-order Runge-Kutta algorithm. 
\cite{differe} The experimental laser pulses along the z-axis are 
estimated to be equivalent to magnetic fields with an amplitude of 0.3 T and a 
full-width at half-maximum of about $200$ fs. \cite{kimel1} 
Therefore, we consider Gaussian fields of the form given in Eq.(\ref{e4a}), with 
$F_0=\sqrt{\pi} \tau_p Amp(H)$, $\tau_p=200$ fs and $Amp(H)$ typically 3000 Oe. 
This value of the effective field is small when compared with the exchange fields, so 
that effects due to the nonlinearity of the equations are negligible. Since the time 
width of the pulse is small on the scale of the mode periods, the results are not 
strongly dependent on $\tau_p$. The linear solution with a delta pulse represents a good 
approximation, provides the right orders of magnitude  but does not reproduce 
exactly the numerical results. Therefore, in all the following figures 
we plot the results obtained by the numerical integration.     

In the first and second subsections we discuss the effects of a Gaussian pulse along 
z and x directions, respectively, when the system at equilibrium is in the 
$\Gamma_4$ phase. In the final one we briefly analyze the resonance modes and 
magnetization dynamics if the $\Gamma_1$ phase is stable at low temperatures.

\subsection{Pulse along z-direction}
As illustrated in Ref. 12, pulses along the z and -z directions excite the 
quasi-antiferro mode. Using the numerical procedure discussed in the preceding section,
we have calculated the oscillating behavior of the ferromagnetic vector
along z focusing on $\Delta M_Z$ at $T=95$ K. 
As shown in Fig.2(a), the magnetization has a sine-like behavior with weak damping on 
a picosecond scale. In addition, opposite fields give rise to 
out-of-phase oscillations in agreement with the experimental data. 
We notice that after 50 picoseconds the amplitude is about half of the initial one, a 
value that is compatible with results reported in Fig.1 of Ref.12. This suggests that
the damping term associated to magnon-magnon scattering provides a reasonable 
description of the reduction of the amplitude as a function of temperature. 

Comparing the experimental data shown in Fig.1 of Ref.12 with the theoretical results 
shown in Fig.2(a) of this paper, we point out similarities and differences. 
Indeed, on the scale of the initial pulse width, there is a strong enhancement of the 
experimental Faraday rotation, probably due to the interference between pump and probe 
pulses. After this transient the oscillations induced by pulses with opposite 
helicities show similar amplitudes in time. In contrast with 
theoretical results, the equilibrium point of the Faraday rotations shows a decay 
and its behavior is different for left- and right-handed polarized pulses. Only later 
the oscillations tend toward a common equilibrium point that is different, however, 
from that before the pulse excitation. 
If the shift of the equilibrium positions is caused by a change of the anisotropy 
constants due to an intrinsic photomagnetic effect, then the amplitude of the 
oscillations should be different for the two helicities, as reported in a recent 
experiment. \cite{hanst} Since this is not the case in dysprosium orthoferrites, 
the change in the orientation equilibrium is more probably associated with a small but 
unavoidable laser-heating effect. Actually, within the model proposed in this paper, a 
shift of the equilibrium configuration could in principle be caused by the magnetic 
pulse during the excitation. \cite{hermann}
This shift is negligible, however, since the amplitude of the effective magnetic field 
is much smaller than the antisymmetrical exchange field $D$. Only by enhancing the 
amplitude of the pulse to values comparable to $D$, the shift becomes important, and, 
upon further increase, even a precessional switching can occur. These 
amplitudes imply very high light intensities that are probably above the damage 
threshold of the samples.

In Fig. 2(b) we focus on the related quantity $d (\Delta M_Z)$, defined as
\begin{equation}
d (\Delta M_Z)=\Delta M_Z (H+)-\Delta M_Z (H-),
\label{e17a}
\end{equation}
with $H+$ and $H-$ indicating positive and negative amplitudes, respectively. 
As reported in the experimental work, 
\cite{kimel1} this quantity is less dependent on initial effects. 
Just after the transient induced by the pulse, this quantity has an amplitude of the 
order of $M_S/16$, in close agreement with the calculated data reported in Fig. 2(b). 
It is worthwhile understanding the order of magnitude of this amplitude by exploiting 
the result of the previous section for the delta pulse. From Eq.(\ref{e14}) we get 
for short times ($t \ll \tau_2$)
\begin{eqnarray}
\Delta {M}_Z \simeq 
2 M_0 R \sin(\gamma F_0) \sin(\omega_{AFM} t), 
\label{e18}
\end{eqnarray}
with $R$ given in Eq.(\ref{e16d}).
The amplitude is determined by the term $R$ depending on the model parameters and by 
$\sin(\gamma F_0)$ related to the pulse intensity. 
Using the experimental data, the impulse $F_0$ is estimated to be of the order of 
$3000 \times 400 $ Oe$\cdot$fs. This implies that $\gamma F_0$ is of the same order as 
the angle $\beta_0$ responsible for the canted antiferromagnetism:
$ \sin(\gamma F_0) \sim \gamma F_0 \sim 2 \beta_0$. Thus, we have
\begin{eqnarray}
\Delta {M}_Z/M_S \simeq 2 R \sin(\omega_{AFM} t).
\label{e19}
\end{eqnarray}
If we consider $R \simeq D/2E$, we obtain the order of magnitude 
$16 d (\Delta {M}_Z)/M_S \sim 16 \times 2 \times 0.02 =0.64 $  not so far from the
experiment. Hence the ratio $D/E$ is fundamental not only for the equilibrium 
configuration, but also for the magnetization dynamics in the canted antiferromagnet.

\begin{figure}[t!]%2
  \begin{center}
  \includegraphics[width=8cm,height=12cm]{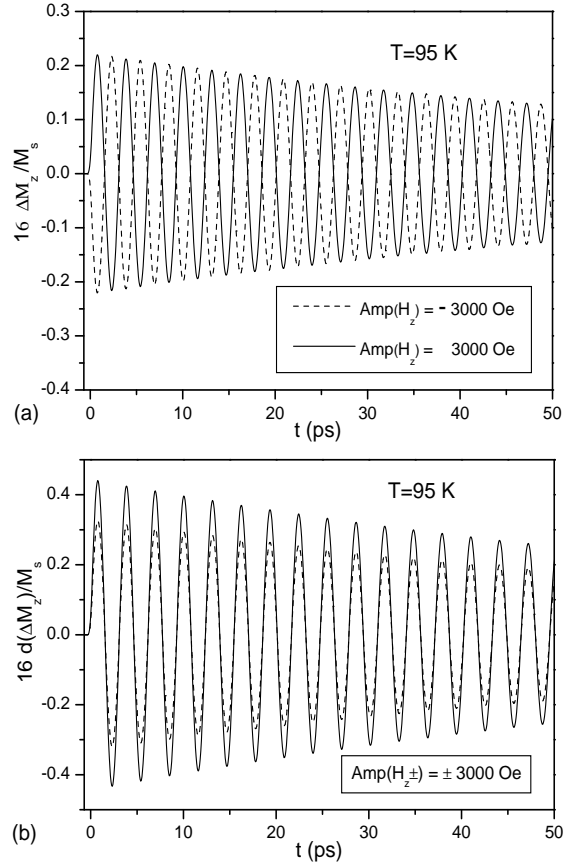}
  \end{center}
%  \vskip-3mm
\caption{
(a) The variation $\Delta {M}_Z$ of the magnetization along z-axis with respect to the 
equilibrium value in units of $M_S/16$ as a function of time for two opposite 
amplitudes.
(b) The difference $ d (\Delta {M}_Z)$ in units of $M_S/16$ between the magnetizations 
obtained upon excitation by opposite fields along z-direction as function of time. The 
dashed line includes the effect of the anisotropy that will be introduced in the next 
subsection. 
}\end{figure}

Within the theoretical approach, the dynamics of the 
antiferromagnetic vectors can be easily obtained, but these quantities have not been 
measured in the experimental work. \cite{kimel1} 
Due to the fact that the equilibrium position corresponds to a maximum of 
$A_X=M_0 (X_1 - X_2)$, the temporal evolution of 
$\Delta A_X=M_0 \Delta \tilde{A}_X$, with $\Delta \tilde{A}_X$ defined in 
Eq.(\ref{e15a}), is characterized by a very small amplitude. From Eq.(\ref{e15}) one can
deduce that the amplitude of $\Delta A_X$ should be at least an order of magnitude 
smaller than that of $\Delta {M}_Z$.  

In Fig. 3 we show the results of the numerical calculation for the quantity 
$\Delta A_Y=M_0 \Delta \tilde{A}_Y$, where $ \Delta \tilde{A}_Y$ is defined in 
Eq.(\ref{e16a}). We point out that now the amplitude is an 
order of magnitude larger than that of $\Delta {M}_Z$ and the response is more sensitive
to the pulse. Actually this is due to the fact that the dynamics of the components along 
the y-axis is strongly influenced by the highest energy scale $E$. \cite{hermann} 
From Eq.(\ref{e16}) we obtain a rough estimate for short times  ($t \ll \tau_2$)  
\begin{equation}
\Delta {A}_Y \simeq 2 M_0 \sin(\gamma F_0) \cos(\omega_{AFM} t).
\label{e20}
\end{equation}
Therefore the amplitude is 
$2 \sin(\gamma F_0) M_0 \sim 2 \beta_0 M_0 \sim 0.022 M_0=M_S$. Finally, we notice 
that, after 50 picoseconds, the damping acts in the same way as for the ferromagnetic 
vector along the z-direction, causing a reduction of about half of the initial 
amplitude.

\begin{figure}[t!]%3
  \begin{center}
  \includegraphics[width=8cm,height=9cm,angle=-90]{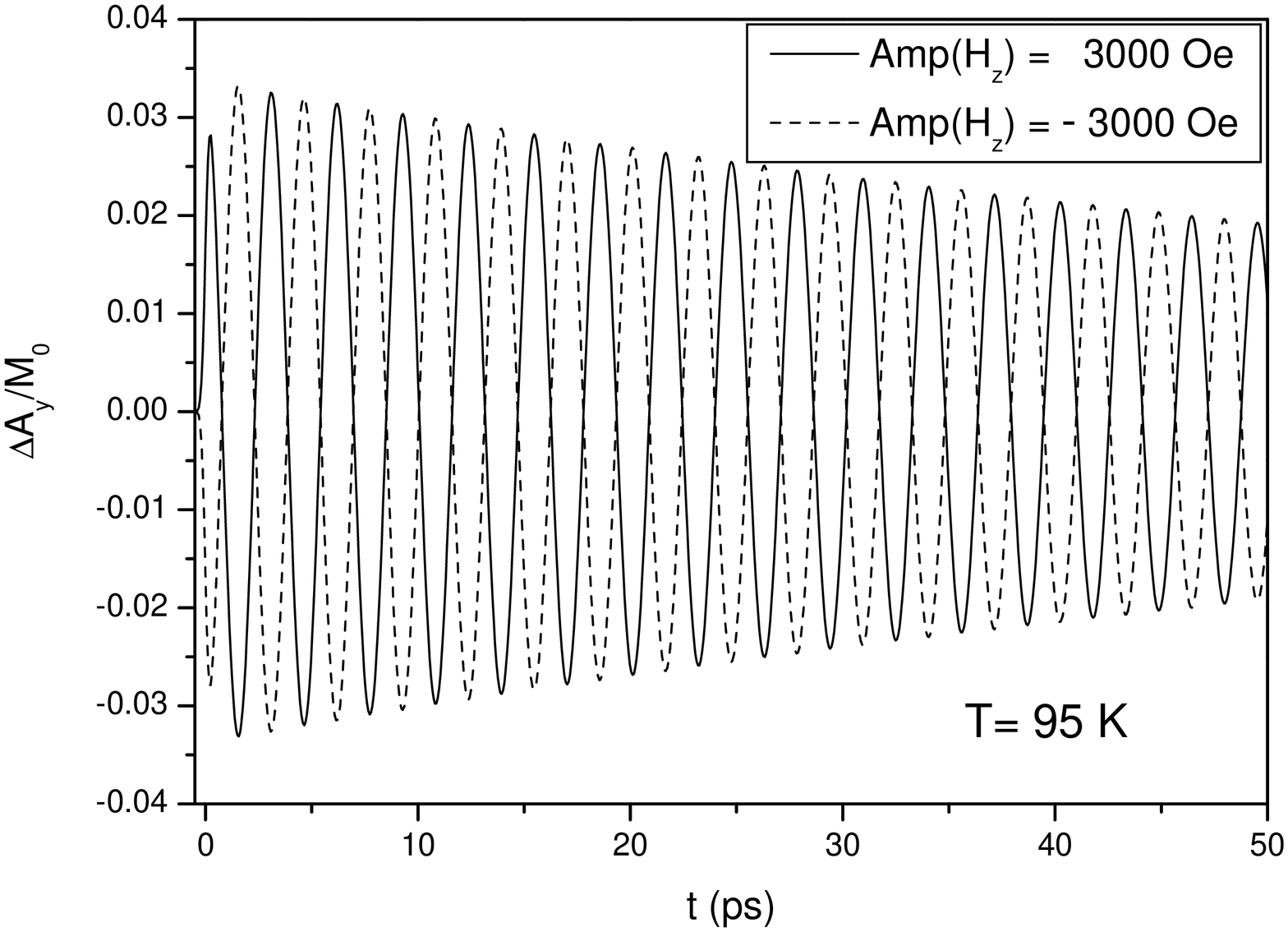}
  \end{center}
   \vskip-3mm
\caption{
The variation of the antiferromagnetic vector along y-axis $\Delta {A}_Y$ with respect 
to $M_0$ as a function of time for two opposite amplitudes.
}\end{figure}

In order to make contact to the experimental measurements, it is interesting to analyze 
the behavior of the magnetization dynamics at different temperatures. As reported in 
Fig.3 of the paper by Kimel {\it et al.},\cite{kimel1} upon heating the sample, 
the frequency of the oscillations increases and the amplitude decreases. The 
oscillations have temperature-dependent frequencies that are very close to those 
measured by Raman experiments at temperatures higher than 50 K. 
In Fig.4 of this paper the calculated $d (\Delta M_Z)$ is shown. 
We find agreement with experimental decay of amplitudes and behavior of the 
frequency upon increasing the temperature for times after the transient. 
Within the theoretical approach, the temperature dependence of the anisotropy constants 
$A_{xx}$ and $A_{zz}$ affects the frequency of the modes, but is not so important in the 
change of the amplitude after the laser transient. The decrease of the magnetization 
seen at a fixed time (of the order of tens of picoseconds) for different temperatures 
is dominated by the increase of the damping constant $\alpha$ in temperature. 
Due to the agreement with experiment, the estimate of the damping constant 
derived on the basis of the magnon-magnon scattering is reliable.

The results of Fig.4 show the magnetization normalized by the sublattice magnetization 
modulus $M_0$. We point out that, in principle, $M_0$ can vary as function of 
temperature. However, the inclusion of the temperature dependence of 
$M_0=M_z/(2 \sin(\beta_0)$) is not easy, since only the magnetization $M_z$ is 
typically experimentally available. 
In any case, even if the angle $\beta_0$ is assumed fixed, the change of $M_z$ is small
in the considered temperature range. \cite{wolf}

\begin{figure}[t!]%4
  \begin{center}
  \includegraphics[width=8cm,height=9cm,angle=-90]{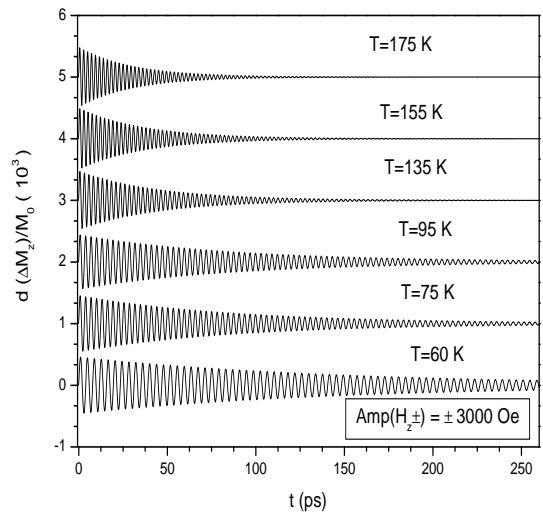}
  \end{center}
   \vskip-3mm
\caption{The difference between the magnetizations along z for
opposite fields as a function of temperature. Every new curve is shifted from the 
previous one along the vertical axis over 1.
}\end{figure}

Before closing this section, we focus on the dependence of the magnetic response
on the amplitude of the pulse. As inferred by Eq.(\ref{e18}), for the effective magnetic
pulses used in the experiments, the ferromagnetic vector is proportional to $F_0$, 
{\it i.e.}, to the intensity of the light pulse used in the inverse Faraday effect. In 
Fig.5 we plot the numerical results for the amplitude of the magnetization  
at t=40 ps, {\it i.e.}, after about 10 periods as in the experimental data. 
\cite{kimelp} 
We find the expected linear behavior as a function of the amplitude of the pulse. 
The value obtained at 5 T for $T=60$ K is not far from unity.
Thus, the saturation of the magnetization along z is reached for this high intensity, 
in agreement with the extrapolation of the experimental data in the inset of Fig.2 of 
Ref.12. 

\begin{figure}[t!]%5
  \begin{center}
  \includegraphics[width=7cm,height=8cm,angle=-90]{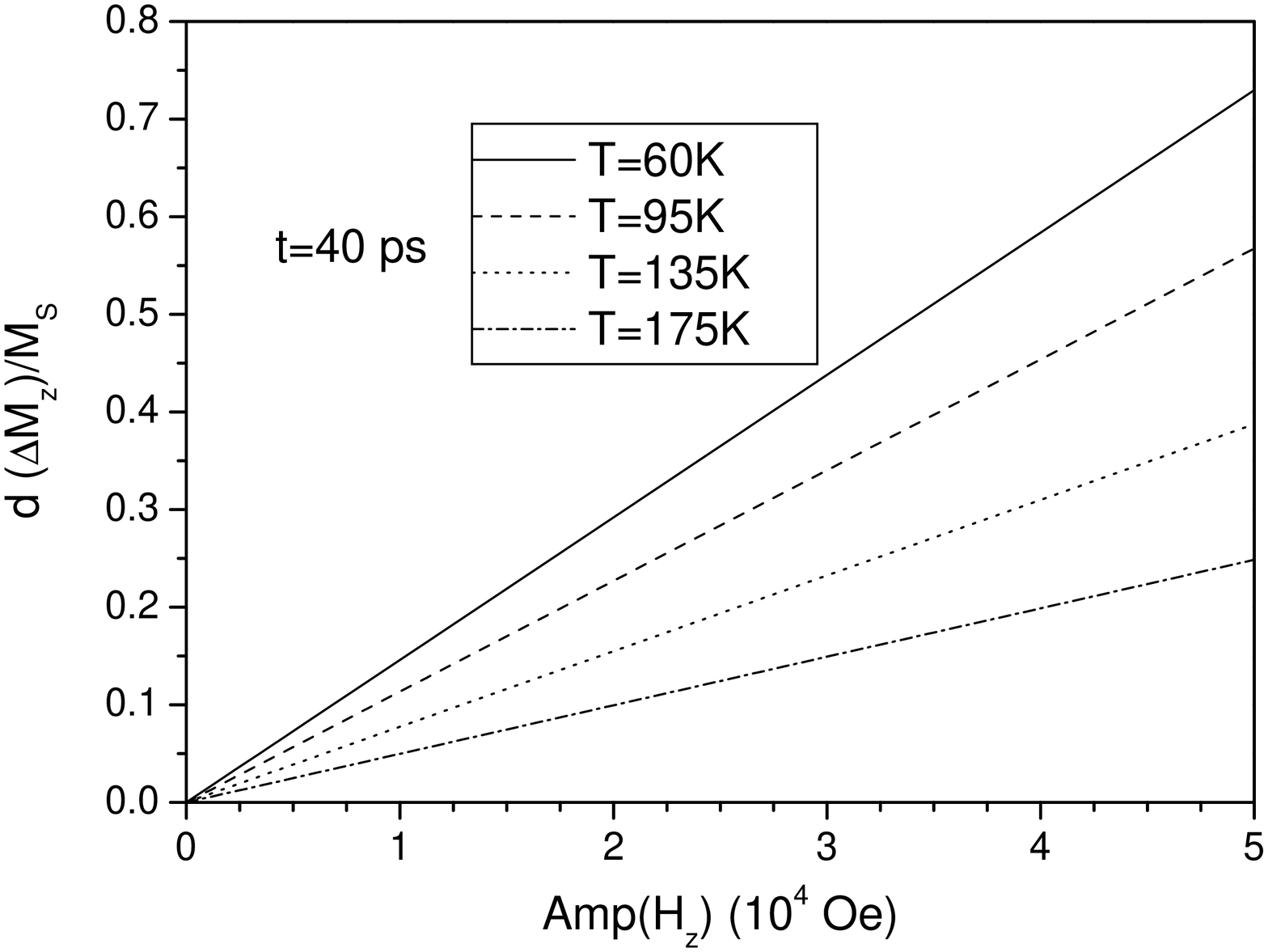}
  \end{center}
   \vskip-3mm
\caption{Amplitude of the spin oscillations as a function of the
amplitude of the magnetic field pulse for different temperatures.
}\end{figure}
Summarizing, in this subsection we have focused on the magnetization dynamics in the 
$\Gamma_4$ phase when the field pulse is directed along the z-axis. We have analyzed 
the amplitude and the decay of the oscillations as a function of temperature finding 
agreement with experimental data for times longer than the initial transient. Actually, 
on the time scale of the initial pulse width, discrepancies with experiments appear. 
These could be due to a small but unavoidable laser-heating effect that is not included
in the present theoretical approach.

\subsection{Pulse along x-direction: role of anisotropy}
In this subsection we analyze the effects of the excitation due to the pulse along the 
x-axis. 

The fields along the x-direction excite another mode. As shown
in Eq.(\ref{e17}), $\Delta \tilde{M}_X$ oscillates with the frequency of the 
quasi-ferro mode in agreement with experiment. On the other hand, the magnitude of the 
magnetic response along x is of the same order as that obtained along the z-axis {\it 
with a pulse of equal amplitude}. 
Even if the torque exerted by the field along x is tiny, 
the subsequent temporal evolution, in particular, that of the variables $T_1$ and 
$T_2$, is able to give a non-negligible amplitude to the response along $x$. 
This is in discrepancy with the experimental results which show that 
the {\it Faraday rotation} along x is at least an order of magnitude smaller 
than that along the z-axis. \cite{kimel1}
Also, as shown in Fig.3 of Ref.12, it does not show any temperature dependence 
following the behavior of the frequency of the quasi-ferro mode, in striking contrast 
with the case of a pulse along the z-axis. 

In order to properly compare the experimental and theoretical results, we should take 
into account several effects in orthoferrites. \cite{simon} 
First of all, the experimental estimate 
of the effective magnetic field pulse due to the inverse Faraday effect is available 
only when the pulse is along the z-axis. Moreover, it is well known that in rare-earth 
orthoferrites optical birefringence is not negligible. \cite{tabor,usachev} Therefore 
the experimental Faraday rotations along the z- and x-axis can be different, even if 
the oscillations of the magnetizations are of the same order. Since the Faraday 
rotation along the x direction is very small in comparison with that along the z-axis, 
part of the effect could be also associated with the anisotropy of the magneto-optical
susceptibility. Indeed, the effective magnetic field generated via the inverse Faraday 
effect can be different for the two orthogonal orientations. Finally, there is another 
source of anisotropy: the field along the z-axis can show a renormalization of the 
coupling to the system that is different from that along the x-axis.   

In order to elucidate this last point, we have included in our model the anisotropy 
induced by the dysprosium ions. As a result, a modification of 
the coupling of the magnetic field to the iron ions takes place. According to a simple 
scheme proposed long ago by Zvezdin and Matveev, \cite{zvezdin} 
the coupling to the field along the z-direction is reduced by the factor 
$1+\eta_z \simeq 0.74$ (see in Fig.2(b) the effect of this reduction). 
Instead, the reduction $1+\eta_x$ of the coupling along the x-axis is temperature 
dependent and can be extracted by the measurements of magnetic properties. 
\cite{zvezdin} With temperature decreasing toward 50 K, these coupling factors can 
become a non-negligible source of anisotropy. As shown in Fig. 6, the difference in 
the amplitudes for the response along the z- and x-axis can become relevant at those 
temperatures. However, as shown in the inset of Fig. 6, the ratio between the 
amplitudes along x and z axis increases with temperature up to an inversion point. 
Therefore the source of the anisotropy associated only to dysprosium ions seems not 
to be effective for the interpretation of the experimental data.

\begin{figure}[t!]%6
  \begin{center}
  \includegraphics[width=8.0cm,height=9.0cm,angle=-90]{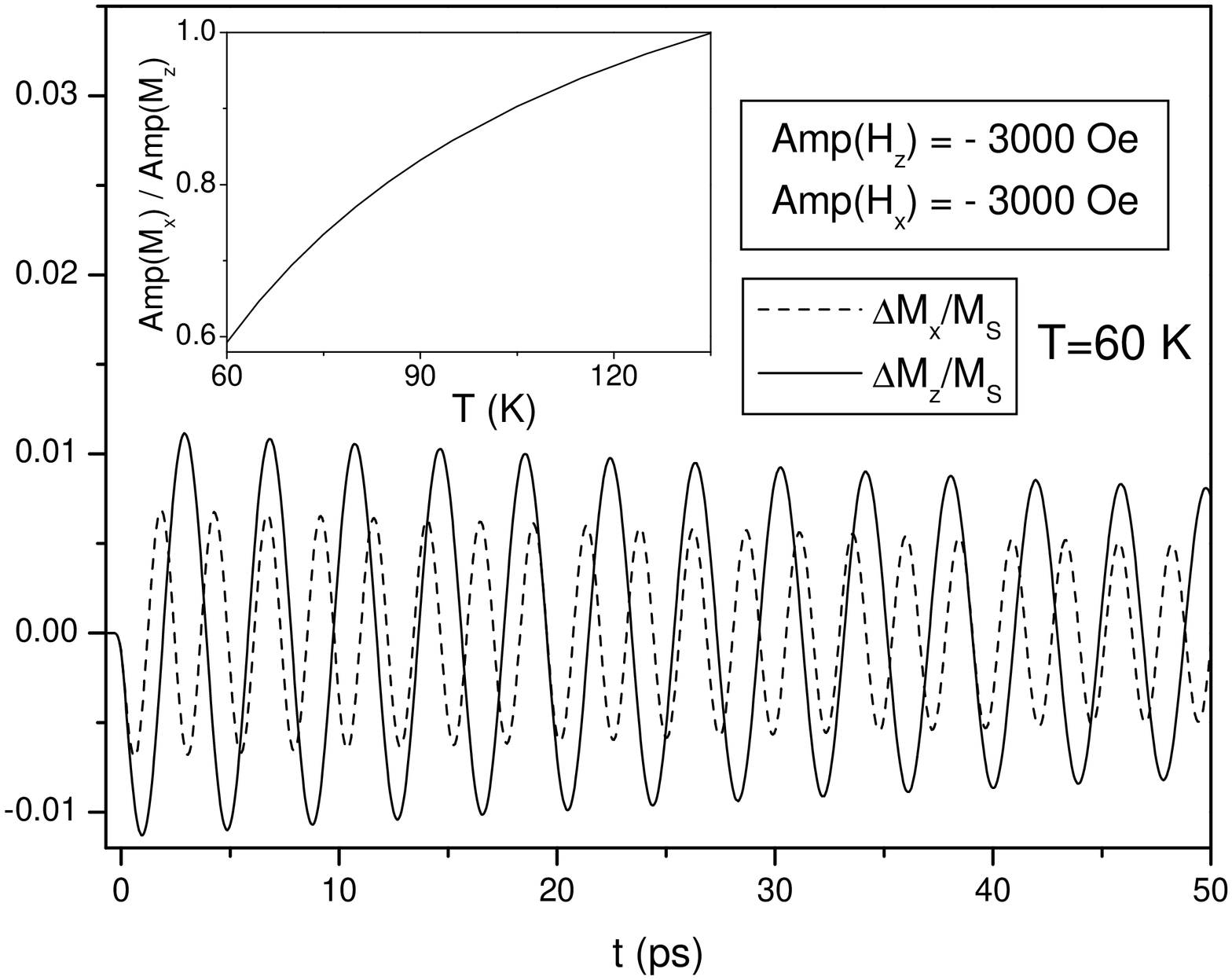}
  \end{center}
%   \vskip-3mm
\caption{
Comparison between amplitudes for excitations with fields along z and x axis as 
function of time at T=60 K. In the inset the ratio of these amplitudes as function of the
temperature for times just after the pulse.  
}\end{figure}

\subsection{Stable phase at low temperatures}
In this last subsection we focus on the $\Gamma_1$ phase that is stable at low 
temperatures. 

The measurements reported in Ref.12 show the maximum value of the amplitude of the 
magnetization between 20 and 50 K. Moreover, the frequency of the photoinduced 
magnetization is constant in this temperature range at the value close to 50 K 
that is characteristic of the $\Gamma_4$ phase. Therefore, the laser pulse is inducing 
excitation modes of the phase that should be unstable at low temperatures.

In order to stabilize the $\Gamma_1$ phase, the free energy (\ref{e0}) has to be 
supplemented with a quartic term, \cite{zvezdin} for instance, a term that decreases the
energy when the moments are antiferromagnetically aligned along the y axis (see Fig.1).
We assume $V \rightarrow V^{\prime}=V-A^{(2)}_{yy} (Y_1-Y_2)^4$ in order to describe 
the system in simple terms.
Upon linearization around the equilibrium configuration $Y_1^{eq}=1$ and $Y_2^{eq}=-1$, 
the two modes characteristic of this phase can be derived. The first one, 
with frequency 
\begin{equation}
\frac{\omega^2_{z}}{\gamma^2}=2 E (32 A^{(2)}_{yy}-2 A_{xx})-D^2,
\label{e5f}
\end{equation}
corresponds to a dynamic with $\Delta X_1=-\Delta X_2$ and $\Delta Z_1=\Delta Z_2$, 
while the second mode, with frequency
\begin{equation}
\frac{\omega^2_{x}}{\gamma^2}=2 E (2 A_{zz}-32 A^{(2)}_{yy})-D^2,
\label{e6f}
\end{equation}
is characterized by $\Delta X_1=\Delta X_2$ and $\Delta Z_1=-\Delta Z_2$. We follow the
procedure of Sec. II: the anisotropy constants are derived by Eqs.(\ref{e5f},\ref{e6f}) 
and the experimental Raman spectra \cite{koshi} by imposing $A^{(2)}_{yy}=0$. At T=50 
we find  $A_{xx}(\Gamma_1)= - 1040$ Oe that is close to $A_{xx}(\Gamma_4)=-640$ Oe 
obtained if the $\Gamma_4$ phase is stable at that temperature.

The values of $A^{(2)}_{yy}$ required to stabilize the $\Gamma_1$ phase are small. 
The free energy corresponding to the equilibrium position 
of the $\Gamma_1$ phase is $V_1=-E-16 A^{(2)}_{yy}$, while that of the $\Gamma_4$ phase 
is at first order in the canting angle 
$\beta_0$ 
\begin{equation}
V_4= -E -D (2 \beta_0) -2 A_{xx}= -E -\frac{D^2}{E} -2 A_{xx}.  
\end{equation}
The condition $V_1<V_4$ implies that 
\begin{equation}
A^{(2)}_{yy}> \frac{D^2}{16 E}+\frac{ A_{xx} }{8}. 
\end{equation}
Taking into account the values of the exchange fields given in the previous section and 
the anisotropy constant $A_{xx}(\Gamma_1)$, we derive $A^{(2)}_{yy}> 60$ Oe. Therefore 
the value of the constant  $A^{(2)}_{yy}$ is consistent with the 
expansion of the free energy being only a fraction of the anisotropy energies 
$A_{xx}$ and $A_{zz}$. 

If the parameter  $A^{(2)}_{yy}$ is of the order of hundreds of Oersted, the phases 
$\Gamma_4$ and $\Gamma_1$ are close in energy. Due to a small light absorption, the 
laser pulse could affect the stability of the system by favoring the $\Gamma_4$ phase 
on a short time scale. 
Therefore the femtosecond pulse could induce a reorientational phase transition 
from $\Gamma_1$ to $\Gamma_4$ state in analogy with the 
antiferromagnetic-to-ferromagnetic phase transition induced by heating with a laser in 
FeRh films. \cite{thiele} 
Finally, we point out that a static magnetic field gives rise to a 
spin reorentational transition in orthoferrites. \cite{zvezdin} Hence the role of the 
magnetic field pulse induced via the inverse Faraday effect could be investigated in 
relation to the perturbation of the phase stability. This is left for future 
investigations.

\section{Summary}
Stimulated by recent experimental results showing ultrafast non-thermal control of 
magnetization by instantaneous photomagnetic pulses in dysprosium orthoferrites, 
a theoretical study of magnetization dynamics has been presented in this paper.  
We have employed a general form of free energy suitable for dysprosium orthoferrites 
whose parameters are derived from experimental measurements. 
We have solved coupled sublattice Landau-Lifshitz-Gilbert equations whose damping 
parameter is determined by considering the scattering rate due to magnon-magnon 
interaction.
Due to the inverse Faraday effect, the magnetic fields perturbing the equilibrium 
configuration can be modeled as Gaussian pulses with amplitude proportional to the 
intensity of the light pulse and time width of the order of hundred femtoseconds. 
The non-linear dynamical equations have been integrated through an optimized Runge-Kutta
algorithm and an analytical solution of the linearized system has been discussed in the 
case when the magnetic field pulse is assumed to have the shape of a delta function. 
This solution provides the right orders of magnitude allowing to interpret the 
experimental results in simple terms. 

We have found that the quasi-antiferro mode is excited by the pulse along the z-axis
and the oscillations of the magnetization have amplitudes compatible with experiment. 
Magnetic fields in opposite directions give rise to out-of-phase oscillations 
showing a behavior in agreement with experimental results for times longer than the 
initial transient. 
We have stressed that the magnetization dynamics is not only strongly influenced 
by the amplitude of the magnetic field pulse, but also by the parameters determining 
the free energy, in particular the ratio between the antisymmetric and symmetric 
exchange energies. The temperature dependence of the magnetization dynamics has been 
discussed showing that the proposed damping mechanism based on magnon-magnon 
scattering can be effective on the picosecond scale. When the field pulse is 
along the x-axis, the quasi-ferro mode is excited, but there are some discrepancies 
in the comparison between theory and data. We point out that the response along the 
x-axis can be strongly influenced in orthoferrites by several effects such as the 
optical birefringence and the anisotropy of the magneto-optical susceptibility.  
Finally, the behavior of the magnetization has been analyzed in the low-temperature 
range where, due to an unavoidable heating effect, the laser pulse could perturb 
the stability between $\Gamma_1$ and $\Gamma_4$ state.

We notice that the model proposed in this work has neglected dipolar contributions 
because they are orders of magnitude smaller than exchange and anisotropy terms. 
Moreover, due to the fact that the effective magnetic field obtained through the 
inverse Faraday effect shows spatial variations negligible on the microscopic scale, 
only the spin-wave modes at zero wave-vector are excited. Since the static 
magnetization changes slowly in the investigated temperature range, the presence of 
spin-waves with wave-vectors different from zero should not provide sizable 
contributions to the dynamic behavior. Therefore, the macrospin approximation employed 
in this paper can be considered reliable.       

Finally, we point out that the approach employed for dysprosium orthoferrites can be 
also generalized to describe the magnetization dynamics of other rare-earth 
orthoferrites, at least in the $\Gamma_4$ phase. The anisotropy constants are the only 
quantities strongly dependent on the rare-earth ion, but these do not play a major role 
in affecting statics and dynamics. On the other hand, the most important values of the 
exchange fields are of the same order in several rare-earth orthoferrites. \cite{koshi}
Clearly the approach proposed in this paper is suitable for magnetic dielectrics 
and not for metallic itinerant magnets. However, up to now, due to the unavoidable light 
absorption, it has been impossible to ascertain the role played by the inverse Faraday 
effect on the magnetization dynamics of itinerant magnets.
 
\section*{Acknowledgments}
We like to thank A.V. Kimel and A. Bringer for critical reading of the manuscript, 
and A. Kirilyuk and S. Woodford for useful discussions.
C.A. Perroni acknowledges financial support by the European RTN network DYNAMICS.


\begin{thebibliography}{99}
\bibitem{notarot}
    We have checked that the circularly polarized magnetic field associated to the 
electromagnetic pulse with frequency corresponding to $1.55$ eV gives a contribution to 
the amplitude of the response that is at least four orders of magnitude less than that 
obtained when the inverse Faraday effect is active.
\end{thebibliography}
 \end{document}